\documentstyle[aps,preprint]{revtex}
\begin{document}
\draft
\title{New explanation of the GAMS results on the $f_0(980)$
production in the reaction $\pi^-p\to \pi^0\pi^0n$ 
\thanks{Dedicated to the memory of Yu.D. Prokoshkin}}
\author{N.N. Achasov and G.N. Shestakov}
\address{Laboratory of Theoretical Physics,\\
S.L. Sobolev Institute for Mathematics,\\
630090, Novosibirsk 90, Russia}
\maketitle
\begin{abstract}
The observed alteration of the S-wave $\pi^0\pi^0$ mass spectrum in the
reaction $\pi^-p\to\pi^0\pi^0n$ with increasing $-t$, i.e., the
disappearance of a dip and the appearance of a peak in the region of the
$f_0(980)$ resonance as $-t$ increases, is explained by the contribution of the
$\pi^-p\to f_0(980)n$ reaction amplitude with the quantum numbers of the $a_1$
Regge pole in the $t$ channel. It is very interesting that nontrivial
evidence for the $a_1$ exchange mechanism in the reaction $\pi^-p\to
\pi^0\pi^0n$ follows for the first time from the experiment on an unpolarized
target. The explanation of the GAMS results suggested by us is compared with
that reported previously. Two ways of experimentally testing these
explanations are pointed out.
\end{abstract} \vspace{0.5cm}

\newpage
\section{Introduction}

Recently, the GAMS Collaboration has continued the investigation of the
reaction $\pi^-p\to\pi^0\pi^0n$ at $P^{\pi^-}_{lab}=38$ GeV [1]. The goal of
the new experiment is to study the $t$ behavior of the S-wave $\pi^0\pi^0$
mass spectrum in the region of the $f_0(980)$ resonance ($t$ is the
square of the four-momentum transferred from the incoming $\pi^-$ to the
outgoing $\pi^0\pi^0$ system). The partial wave analysis performed in the
range $0<-t<1$ GeV$^2$ gave a very interesting and unexpected result.
The $f_0(980)$ resonance has been seen as a dip in the
S-wave $\pi^0\pi^0$ mass spectrum for
$-t<0.2$ GeV$^2$ (see Fig. 1a), where the reaction $\pi^-p\to\pi^0\pi
^0n$ is dominated by the one-pion exchange mechanism, \footnote{As is well
known, such a manifestation of the $f_0(980)$ resonance, due to its strong
destructive interference with the background, was observed in a large number 
of previous experiments on the reactions $\pi N\to\pi\pi N$ and $\pi N\to\pi
\pi\Delta(1232)$, and according to their results, it has also been well
established in the reaction $\pi\pi\to\pi\pi$ (see, for example, Refs. [2-11],
and for reviews, Refs. [12-15]).} whereas for $-t>0.3$ GeV$^2$, it has
been observed as a distinct peak (see Figs. 1b-f). This dip and peak behavior
of the $f_0(980)$ has also been seen in the Brookhaven experiment on the
reaction $\pi^-p\to\pi^0\pi^0n$ at $P^{\pi^-}_{lab}=18$ GeV [16]. A partial
wave analysis of these data is presently being undertaken [16].

In this work we show that the observed alteration of the S-wave $\pi^0\pi^0$
mass spectrum in the reaction $\pi^-p\to\pi^0\pi^0n$ with increasing $-t$ can
be explained by the contribution of the $\pi^-p\to f_0(980)n$ reaction
amplitude with quantum numbers of the $a_1$ Regge pole in the $t$ channel. So
far this amplitude has been very poorly studied experimentally.

In fact, we suggest the following plausible scenario. At small $-t$, the 
reaction $\pi^-p\to(\pi^0\pi^0)_S\,n$ is dominated by the one-pion exchange 
mechanism, and the $f_0(980)$ resonance manifests itself in the $(\pi^0\pi^0
)_S$ mass spectrum as a minimum ($(\pi
\pi)_S$ denotes a $\pi\pi$ system with the orbital angular momentum $L=0$).
However, the one-pion exchange contribution decreases very rapidly with $-t$
(as is known, at least $85-90\%$ of the one-pion exchange cross section for
the reactions $\pi N\to\pi\pi N$ originate from the region $-t<0.2$ GeV$^2$).
The most remarkable fact is that the reactions $\pi N
\to(\pi\pi)_S\,N$ at high energies involve only two types of $t
$-channel exchanges, namely, those with quantum numbers of the $\pi$ and $a_1$
Regge poles. Thus, it is very probable that the reaction $\pi^-p\to(\pi
^0\pi^0)_S\,n$ at large $-t$ is dominated by the $a_1$ exchange, and that the
$f_0(980)$ resonance produced by this mechanism shows itself as a peak. Notice
that a similar manifestation of the $f_0(980)$ resonance has been
observed in many reactions not involving $\pi$ exchange (i.e., in which the $
\pi\pi$ interaction in the initial state is absent). For example, the $f_0(980)
$ resonance has been seen as a clear peak in the two-pion mass spectra in the
reaction $\pi^-p\to\pi^0\pi^0n$ near threshold and for $-t$ from 0.33 to 0.83
GeV$^2$, where the one-pion exchange is small [17], in the reaction
$K^-p\to\pi^+\pi^-(\Lambda,\Sigma)$ at 13 GeV [18], in the $J/\psi\to\phi
\pi^+\pi^-$ [19] and $D_s^+\to\pi^+\pi^+\pi^-$ [20] decays, in the reaction
$\gamma\gamma\to\pi^0\pi^0$ [21], and also in the inclusive $\pi^+\pi^-$
production in $\gamma p$, $\pi^\pm p$, $K^\pm p$ [22], and $e^+e^-$ [23]
collisions.

Our explanation of the GAMS results may be unambiguously verified
experimentally in the reactions $\pi N\to\pi\pi N$ on polarized targets because
this makes possible direct measurements of the interference between the $\pi
$ and $a_1$ exchange amplitudes. In a cross section summed over the nucleon
polarizations, the contributions of these amplitudes are noncoherent and,
generally speaking, they cannot be separated without additional
assumptions. It is interesting to note in this connection that the GAMS
Collaboration has probably become the first who succeeded in discovering a
nontrivial evidence for the $a_1$ exchange mechanism in the reaction $\pi^-p
\to(\pi^0\pi^0)_S\,n$ on an unpolarized target.
\footnote{As is known, the results of the measurements of the reactions $\pi^
\pm N_\uparrow\to\pi^+\pi^-N$ on polarized targets are indicative of the $a_1$
exchange mechanism most definitely in the case of the $\rho^0(770)$
production [24-26]. However, in the $\pi\pi$ invariant mass region around 1
GeV, rather large experimental uncertainties in the available data present
considerable problems for certain conclusions. Nevertheless, in a new analysis
of the $\pi^-p_\uparrow\to\pi^+\pi^-n$ data at 17.2 GeV [6,25], which has been
performed very recently in Ref. [27], one emphasizes
that the $a_1$ exchange amplitude cannot be neglected especially
around 1 and 1.5 GeV.}

In Sec. II, we perform a simultaneous description of the GAMS data on the
reaction $\pi^-p\to(\pi^0\pi^0)_S\,n$ [1] and the CERN-Munich data on the
S-wave $\pi\pi$ scattering in an invariant mass region around 1 GeV [5]. We
consider three simple parametrizations of the S-wave $\pi\pi\to\pi\pi$ reaction
amplitude. As to the corresponding amplitude of the reaction $\pi^*\pi\to\pi\pi
$ (where $\pi^*$ denotes a Reggeized pion), it is constructed by using the
$t$ dependence factorization assumption which was
extensively applied previously to obtain the $\pi\pi$ scattering data
(see, for example, Refs. [5,7,8,27-29]). In parametrizing the $\pi^-p\to(\pi^0
\pi^0)_S\,n$ reaction amplitude due to the $a_1$ exchange, we use the above
qualitative reason
based on the observations of the $f_0(980)$ resonance in the reactions not
involving $\pi$ exchange. All considered parametrizations of the
$\pi^-p\to(\pi^0\pi^0)_S\,n$ reaction amplitudes give similar
results and, on the whole, quite reasonable fits to the GAMS data. In Sec. III,
we compare our explanation of the GAMS data with that reported previously in
Ref. [30] and point out two direct ways to test these explanations.
The explanation of Ref. [30] differs crucially from ours in that it is based
entirely on one-pion exchange or exchanges with these quantum numbers. Such a 
restriction, as we show, leads, in particular, to rather exotic predictions 
for the $t$ distributions of the $\pi^-p\to(\pi^0\pi^0)_S\,n$ events. Our 
conclusions are briefly summarized in Sec. IV.

\section{Alteration of the $(\pi^0\pi^0)_S$ mass spectrum in the
$\lowercase{f_0(980)}$ region in the reaction $\pi^-\lowercase{p}\to(
\pi^0\pi^0)_S\,\lowercase{n}$}

We shall consider the reaction $\pi^-p\to(\pi^0\pi^0)_S\,n$ within the
framework of the simplest Regge pole model and write the unpolarized
differential distribution of the $\pi^-p\to(\pi^0\pi^0)_S\,n$ events
at fixed $P^{\pi^-}_{lab}$ in the following form:
\begin{eqnarray} \frac{d^2N}{dmdt}=
\left|\,A_\pi\ \frac{\sqrt{-t}}{t-m^2_\pi}\ e^{\tilde b_\pi(t-m^2_\pi)/2}\ e^{
-i\pi\alpha_\pi(t)/2}\ \,\sqrt{m/\rho_{\pi\pi}}\ \,T_{\pi^*\pi\to\pi\pi}(m,t)\,
\right|^2+\nonumber\\[3pt] +\left|\,A_{a_1}\ (1+t\,C)\ e^{\tilde b_{a_1}t/2}\
i\ e^{-i\pi\alpha_{a_1}(t)/2}\ \,\sqrt{m}\ \,R_{a_1\pi\to\pi\pi}(m,t)\,\right|^
2\ .\,\ \ \ \ \end{eqnarray}
Here the first and second terms correspond to the $\pi$ and $a_1$ Regge pole
contributions, respectively (the $\pi$ and $a_1$ exchanges do not interfere 
because, at high energies, they contribute to different helicity amplitudes),
$\alpha_\pi(t)=\alpha'_\pi(t-m^2_\pi)$ and
$\alpha_{a_1}(t)=\alpha_{a_1}(0)+\alpha'_{a_1}t$ are the trajectories of these
poles, $m$ is the invariant mass of the final $\pi\pi$ system, $A_\pi$ and $A_
{a_1}$ are the normalization constants, $T_{\pi^*\pi\to\pi\pi}(m,t)$ and $R_{a
_1\pi\to\pi\pi}(m,t)$ are the S-wave amplitudes for the subprocesses $\pi^{*+}
\pi^-\to\pi^0\pi^0$ and $a_1^+\pi^-\to\pi^0\pi^0$, respectively,
$\ \rho_{\pi\pi}=(1-4m^2_\pi/m^2)^{1/2}$, the slope $\tilde b_\pi=2\alpha'_\pi
\ln(P^{\pi^-}_{lab}/1$GeV$)+b_{\pi NN}$, i.e., it incorporates the slope of the
Reggeized pion propagator and the slope of the $\pi^* NN$ residue taken in the
exponential form, and the slope $\tilde b_{a_1}$ has a similar structure.
According to the physical reasons which were discussed in the
literature, the $a_1$ Regge pole amplitude has to have the so-called
sense-nonsense wrong signature zero at $\alpha_{a_1}(t=t_0)=0$, and hence, to
be proportional to $\alpha_{a_1}(t)$ (see, for example, Refs. [31-33]). Thus,
the factor $(1+t\,C)$ in the second term of Eq. (1) can be understood as the
ratio $\alpha_{a_1}(t)/\alpha_{a_1}(0)=1+t\,\alpha'_{a_1}/\alpha_{a_1}(0)$.
However, the value of $\alpha'_{a_1}/\alpha_{a_1}(0)$ is in fact unknown
[32,33], and therefore, we consider $C$ as a free parameter.
According to isotopic symmetry,
\begin{eqnarray} T_{\pi^*\pi\to\pi\pi}(m,t)=T^0_0(m,t)-T^2_0(m,t)\ ,\qquad
R_{a_1\pi\to\pi\pi}(m,t)=R^0_0(m,t)-R^2_0(m,t)\ ,\end{eqnarray} where $T^I_L(m
,t)$ and $R^I_L(m,t)$ are the amplitudes with $L=0$ and isospin $I=0,2$ for the
subprocesses $\pi^*\pi\to\pi\pi$ and $a_1\pi\to\pi\pi$, respectively; the
amplitude $R^2_0(m,t)$ is assumed negligible.
Now we suppose that the $t$ dependences of the amplitudes
$T^I_L(m,t)$ for the reaction $\pi^*\pi\to\pi\pi$ can be extracted in the form
of overall exponential form factors. Thus we put \begin{eqnarray} T^0_0(m,t)=e^
{b^0_0(t-m^2_\pi)/2}\ T^0_0(m)\ ,\qquad T^2_0(m,t)=e^{b^2_0(t-m^2_\pi)/2}\ T^2_
0(m)\ ,\end{eqnarray} where the amplitudes $T^0_0(m)$ and $T^2_0(m)$ depend
only on $m$ and are determined by the on-mass-shell dynamics of the $\pi\pi$
scattering. This assumption about the $t$ dependence factorization, together
with the concrete shape of this dependence, was widely used as a simple
working tool to obtain the $\pi\pi$ scattering data and gave results which 
were in close agreement with those of the more general Chew-Low extrapolation 
method [3-13,27-29]. \footnote{
For the pronounced solitary $\rho(770)$ and $f_2(1270)$ resonances produced in
the reactions $\pi N\to\pi\pi N$ in the low $-t$ region via the one-pion
exchange, the factorization of the $t$ and $m$ dependences for the
$\pi^*\pi\to\rho(770)\to\pi\pi$ and $\pi^*\pi\to f_2(1270)\to\pi\pi$ amplitudes
is quite natural. However, in the S-wave case, the situation is more
complicated. There are at least two strongly interfering contributions in the
$L=I=0$ $\pi^*\pi\to\pi\pi$ channel at $m\approx1$ GeV, namely, the narrow $f_0
(980)$ resonance and the smooth large background which can be parametrized, for
example, in terms of
a broad elastic $\pi\pi$ resonance [34,35]. Even though the $t$ dependence
factorizes for each contribution, the whole $L=I=0$ $\pi^*\pi\to\pi\pi$
amplitude may possess this property only if the various
contributions have rather close $t$ dependence. In connection with the
GAMS results, we discuss the $L=I=0$ $\pi^*\pi\to\pi\pi$ amplitude in the
region of the $f_0(980)$ resonance beyond the $t$ dependence factorization
assumption at the end of this section and also in Sec. III.} Usually, the
factorization assumption is
applied to the $\pi N\to\pi\pi N$ one-pion exchange amplitudes in the
region $0<-t<(0.15-0.2)$ GeV$^2$ [5,7,8,27,29]. We shall use Eq. (3) as ``a
zeroth approximation" (in the sense of a number of addition assumptions and new
fitted parameters) for all $-t$ of interest from 0 to 1 GeV$^2$. Also we adopt
a similar representation for $t<0$ for the amplitude $R^0_0(m,t)$ of the
subprocess $a_1\pi\to\pi\pi$,\begin{eqnarray}  R^0_0(m,t)=e^{c^0_0t/2}\ R^0_0
(m)\ .\end{eqnarray} Note that some smooth $m$ dependence of the slopes $b^0_0
$, $b^2_0$, and $c^0_0$ is not excluded. However, in the considered relatively
narrow $m$ range near the $f_0(980)$ resonance, $0.8<m<1.1$ GeV, we assume for
simplicity that $b^0_0$, $b^2_0$, and $c^0_0$ are constant. From the fit to
the data [1], the values of the overall slopes of the corresponding amplitudes,
namely, $b^0_\pi=\tilde b_\pi+b^0_0\,,\ $ $b^2_\pi=\tilde b_\pi+b^2_0
\ $, and $\ b^0_{a_1}=\tilde b_{a_1}+c^0_0\ $ will be determined (see
Eqs. $(1)-(4)$).

Let us now turn to the description of the model for the amplitudes $T^0_0(m)$,
$T^2_0(m)$, and $R^0_0(m)$. On the mass shell of the reaction $\pi\pi\to\pi\pi$
\begin{eqnarray} T^0_0(m)=(\eta^0_0\ e^{2i\delta^0_0}-1)/2i \ , \qquad  T^2_0
(m)=(\eta^2_0\ e^{2i\delta^2_0}-1)/2i \ ,\end{eqnarray}
where $\delta^I_L$ and $\eta^I_L$ are the phase shifts and elasticities which
are functions of $m$. The data on the $L=0$, $I=2$ $\pi\pi$ channel in the
region $2m_\pi<m<1.2$ GeV are described very well by $\eta^2_0=1$ and $\delta^
2_0=-0.87q_\pi/(1+0.16q^2_\pi)$, where $\delta^2_0$ is in radians if $q_
\pi=m\rho_{\pi\pi}/2$ is taken in units of GeV (see, for example, Ref. [36]).
At $m\approx1$ GeV, $\delta^2_0\approx-23^\circ$. In the $L=I=0$ $\pi\pi$
channel, a very sharp rise of the phase $\delta^0_0$ near the $K\bar K$
threshold (see Figs. 2a and 3a), together with a sharp drop of the elasticity
$\eta^0_0$ just above the $K\bar K$ threshold (see Figs. 2b and 3b), is usually
interpreted in term of the $f_0(980)$ resonance coupled to the $\pi\pi$ and $K
\bar K$ channels [2-15,37]. However, in the $L=I=0$ $\pi\pi\to\pi\pi$ cross
section this puzzling
state shows itself not as a peak, but as a dip which occurs just below the $K
\bar K$ threshold, and in fact, the cross section vanishes at a
minimum point. Formally, this is because the phase $\delta^0_0$
goes through 180$^\circ$, but not though 90$^\circ$, in the resonance region
and $\eta^0_0=1$ with a good accuracy for $m<2m_K$. Note that the $I=2$
wave admixture shifts a minimum in the $L=0$ $\pi^+\pi^-\to\pi^0\pi^0$ reaction
cross section approximately by 10 MeV to a lower mass region. \footnote
{As is seen from Fig. 1a, the observed $\pi^-p\to(\pi^0\pi^0)_S\,n$ cross
section does not vanish at a minimum but accounts for about $1/3$ of the
cross section at the side maxima. This is mainly because of a finite
experimental $\pi^0\pi^0$ mass resolution which for the GAMS-2000 spectrometer
has been characterized by a Gaussian distribution with the dispersion $\sigma
_m\approx20$ MeV at $m\approx1$ GeV [1]. In the fit to the GAMS data, we
certainly take into account this Gaussian smearing.} Let us write the
amplitudes $T^0_0(m)$ and $R^0_0(m)$ as\begin{eqnarray} T^0_0(m)=
\frac{e^{2i\delta_B}-1}{2i}\ +\ e^{2i\delta_B}\ T_{\pi\pi\to\pi\pi}^{res}(m)\ ,
\qquad R^0_0(m)=e^{i\delta_B}\ R_{a_1\pi\to\pi\pi}^{res}(m)\ ,\end{eqnarray}
where $\delta_B$ is the phase shift due to the smooth elastic background in the
$\pi\pi$ channel, whereas $T_{\pi\pi\to\pi\pi}^{res}(m)$ and $R_{a_1\pi
\to\pi\pi}^{res}(m)$ are the amplitudes due to the contributions of the mixed
inelastic resonances. If we put $T_{\pi\pi\to\pi\pi}^{res}
(m)=(\eta_{res}e^{2i\delta_{res}}-1)/2i$, we find from Eqs. (5) and (6)
that $\delta^0_0=\delta_B+\delta_{res}$ and $\eta^0_0=\eta_{res}$. To
parametrize the resonance contributions we use the so-called propagator method
[14,38,39] and write the amplitude $\tilde T_{ab\to cd}^{res}(m)$ for the
process $ab\to cd$ in the following form (which satisfies the
unitarity condition): \begin{eqnarray}\tilde T_{ab\to cd}^{res}(m)=\sum_{r,r'}
g_{rab}\ G^{-1}_{r\,r'}(m)\ g_{r'cd}\ , \end{eqnarray} where the sum is
evaluated over the resonances $r,\ r'$ ($r\,(r')=r_1,\ r_2,\ ...$),
$\ G_{r\,r'}(m)$ is the inverse propagator matrix for a resonance complex,
\begin{eqnarray} G_{r\,r'}(m)=\left( \begin{array}{ccc}
D_{r_1}(m) & -\Pi_{r_1r_2}(m) & ...\\-\Pi_{r_1r_2}(m) & D_{r_2}(m) & ...\\... &
... & ...\end{array} \right) \ , \end{eqnarray} \begin{eqnarray}
D_r(m)=m^2_r-m^2+Re\Pi_r(m_r)-\Pi_r(m)\ ,\end{eqnarray} $m_r$ and $\,g_{rab}$,
$\,g_{r'cd}$ are, respectively, the masses and the coupling constants of the
unmixed resonances. Since we are interested in a mass region around 1 GeV, we
can restrict ourselves to the simplest case of resonances coupled only to the
$\pi\pi$ and $K\bar K$ decay channels. We also imply that the resonance
production occurs in $\pi\pi$ and $a_1
\pi$ collisions (recall that the $a_1$ means here not a particle but a
Reggeon). Then we can take, in Eq. (9),
\begin{eqnarray}\Pi_r(m)=\sum_{cd=\pi\pi,\,K\bar K}g^2_{rcd}\ \rho_{cd}\left(i+
\frac{1}{\pi}\ln\frac{1-\rho_{cd}}{1+\rho_{cd}}\right) \end{eqnarray} and write
the off-diagonal elements of the matrix $G_{r\,r'}(m)$ (see Eq. (8)),
responsible for the resonance mixing, as \begin{eqnarray}\Pi_{r\,r'}(m)=C_{r\,r
'}\ +\sum_{cd=\pi\pi,\, K\bar K}g_{rcd}\ g_{r'cd}\ \rho_{cd}\left(i+\frac{1}{
\pi}\ln\frac{1-\rho_{cd}}{1+\rho_{cd}}\right)\ , \end{eqnarray} where $C_{r\,r'
}$ are the mixing parameters, $\rho_{K\bar K}=(1-4m_K/m^2)^{1/2}$ \ for $m>2m_K
\ $, and $\ \rho_{K\bar K}\to i|\rho_{K\bar K}|$ in the region $0<m<2m_K$. Here
we neglect the $K^+K^-$ and $K^0\bar K^0$ mass difference and put $m_K=(m_{K^+}
+m_{K^0})/2$. Above the corresponding threshold, the partial decay width of the
resonance $r$ is $\ \Gamma_{rcd}(m)=g^2_{rcd}\ \rho_{cd}/m\ $. Using Eqs. (6)
and (7) with due regard for the normalizations as defined in Eqs. $(1)-(5)$, we
finally obtain
\begin{eqnarray} T^{res}_{\pi\pi\to\pi\pi}(m)=\rho_{\pi\pi}\ \tilde
T^{res}_{\pi\pi\to\pi\pi}(m)\ ,\qquad R^{res}_{a_1\pi\to\pi\pi}(m)=\sqrt{\rho_
{\pi\pi}}\ \tilde T^{res}_{a_1\pi\to\pi\pi}(m)/g_{r_1a_1\pi}\ ,\end{eqnarray}
where the second relation implies, in particular, that the coupling constant
$g_{r_1a_1\pi}$ is taken up by the normalization constant
$A_{a_1}$ in Eq. (1).

Within the framework of the above model, we present the three simplest variants
of the fit to the data [5] on $\delta^0_0$ and $\eta^0_0$ in the $f_0(980)$ 
mass region. In
variant 1, we assume that the amplitude $T^0_0(m)$ (see Eq. (6)) is dominated
by a single resonance and a background, in variant 2 by two mixed resonances
and a background, and in variant 3 by two mixed resonances.

Variant 1 yields the most economical and transparent parametrization. Using
Eqs. $(6)-(10)$ and (12), we find in this case
\begin{eqnarray} T^0_0(m)=\frac{e^{2i\delta_B}-1}{2i}\ +\ e^{2i\delta_B}
\ \frac{m\Gamma_{f_0\pi\pi}(m)}{D_{f_0}(m)}\ , \qquad R^0_0(m)=e^{i\delta_B}\
\frac{\sqrt{m\Gamma_{f_0\pi\pi}(m)}}{D_{f_0}(m)}\ ,\end{eqnarray} where $f_0$
is taken as a suitable notation for a single $r_1$ resonance and the background
phase $\delta_B=a+mb$. The parametrization of $T^0_0(m)$ as given by Eq. (13)
permits us to obtain a good fit to the data on $\delta
^0_0$ and $\eta^0_0$ in the region $0.8<m<1.2$ GeV (see the solid curves in
Figs. 2a,b). The corresponding parameters of the background and resonance
are $\delta_B=35.5^\circ+47^\circ(m/$GeV), $\ m_{f_0}=979$ MeV, $\ g^2_{f_0\pi
\pi}=0.075$ GeV$^2\ $ and $\ g^2_{f_0K\bar K}=0.36$ GeV$^2$. Note that the
above simple representation for $T^0_0(m)$ also was used for a similar
purpose in a set of earlier analyses (see, for example, [3,9,35,40,41]).
It is obvious that in this case a dip in the $L=I=0$ $\pi\pi\to\pi\pi$ reaction
cross section in the $f_0(980)$ resonance region is due to the
destructive interference between the resonance and the background whose
contributions are near the S-wave unitarity limit.

Variant 2 allows a good fit to the data on $\delta^0_0$ to be attained in the
wider $m$ interval from 0.6 to 1.7 GeV (see also Ref. [39]) and also
turns out to be more flexible for the construction of the
$\pi^-p\to(\pi^0\pi^0)_S\,n$ reaction amplitude due to the $a_1$
exchange. In this case, using Eqs. $(6)-(12)$, we have
\begin{eqnarray} T^0_0(m)=\frac{e^{2i\delta_B}-1}{2i}\ +\ e^{2i\delta_B}\,\rho_
{\pi\pi}\,\times \ \ \ \ \ \ \ \ \ \ \ \ \ \ \ \ \ \ \ \ \ \ \ \ \ \ \ \
\nonumber\\[3pt] \times\,\frac{g_{r_1\pi\pi}[D_{r_2}(m)\,g_{r_1\pi\pi}+\Pi_{r_1
r_2}(m)\,g_{r_2\pi\pi}]+g_{r_2\pi\pi}[D_{r_1}(m)\,g_{r_2
\pi\pi}+\Pi_{r_1r_2}(m)\,g_{r_1\pi\pi}]}{D_{r_1}(m)D_{r_2}(m)-\Pi^{\,2}_{r_1
r_2}(m)}\ , \end{eqnarray}
\begin{eqnarray}  R^0_0(m)=e^{i\delta_B}\,\sqrt{\rho_{\pi\pi}}\,
\times\ \ \ \ \ \ \ \ \ \ \ \ \ \ \ \ \ \ \ \ \ \ \ \ \ \ \ \ \ \ \ \ \ \ \ \ \
\nonumber\\[3pt] \times\,\frac{[D_{r_2}(m)\,g_{r_1\pi\pi}+\Pi_{r_1r_2}
(m)\,g_{r_2\pi\pi}]+(g_{r_2a_1\pi}/g_{r_1a_1\pi})[D_{r_1}(m)\,g_{r_2
\pi\pi}+\Pi_{r_1r_2}(m)\,g_{r_1\pi\pi}]}{D_{r_1}(m)D_{r_2}(m)-\Pi^{\,2}_{r_1
r_2}(m)}\ ,\end{eqnarray} where $\delta_B=\rho_{\pi\pi}\,(a+mb)$. In the
following, while referring to this variant, the lighter resonance $r_1$ will
be denoted by $f_0$, and $r_2$ by $\sigma$. The curves shown in Figs. 3a,b are
the result of the fit to the data on $\delta^0_0$ and $\eta^0_0$
using Eq. (14). These curves correspond to the following values of the
parameters: $\ m_{f_0}=0.966$ GeV, \  $g^2_{f_0\pi\pi}=0.09$ GeV$^2$, \
$g^2_{f_0K\bar K}=0.36$ GeV$^2$, \ $m_\sigma=1.58$ GeV, \ $g^2_{\sigma\pi\pi}=0
.73$ GeV$^2$, \ $g^2_{\sigma K\bar K}=0.002$ GeV$^2$, \ $C_{f_0\sigma}=\pm0.37$
GeV$^2$, \ and \ $\delta_B=\rho_{\pi\pi}(3^\circ+50^\circ(m/$GeV)). Note that
$C_{f_0\sigma}$ is defined up to a sign, but in so doing
\ $C_{f_0\sigma}\,g_{f_0\pi\pi}\,g_{\sigma\pi\pi}>0$, \ and \ $g_{f_0\pi\pi
}\,g_{\sigma\pi\pi}\,g_{f_0K\bar K}\,g_{\sigma K\bar K}<0$.

In variant 3, the amplitudes $T^0_0(m)
$ and $R^0_0(m)$ are defined by Eqs. (14) and (15) with $\delta_B=0$. We
consider this variant mainly to ease the following discussion of the
results presented in Ref. [30] (see Sec. III). The fit to the data on $\delta^0_
0$ and $\eta^0_0$ in the region $0.8<m<1.2$ GeV with variant 3 gives
$\ m_{r_1}=0.88$ GeV, \ $g^2_{r_1\pi\pi}=0.45$ GeV$^2$, \
$g^2_{r_1K\bar K}=0.57$ GeV$^2$, \ $m_{r_2}=1.23$ GeV, \ $g^2_{r_2\pi\pi}=0.74$
GeV$^2$, \ $g^2_{r_2K\bar K}=0.09$ GeV$^2$, \ $C_{r_1r_2}=\pm0.67$ GeV$^2$, \
$C_{r_1r_2}\,g_{r_1\pi\pi}\,g_{r_2\pi\pi}>0$ \ and \ $g_{r_1\pi\pi}\,g_{r_2\pi
\pi}\,g_{r_1K\bar K}\,g_{r_2K\bar K}<0\ $ (see the dashed curves in Figs.
2a,b).

Now we use the obtained parameters to describe the GAMS data on the $(\pi^0\pi
^0)_S$ mass spectra in the reaction $\pi^-p\to(\pi^0\pi^0)_S\,n$ which are
shown in Figs. 1a-f. For each of the above variants
we perform the fit to these data using Eq. (1) folded with
a Gaussian mass distribution (see footnote 4) and integrated over $t$ in six
intervals indicated in Figs. 1a-f. For variant 1 we use Eqs. $(2)-(4)$, and
(13), and for variants 2 and 3 Eqs. $(2)-(4)$, (14), and (15). As is seen
from Figs. 1a-f, the observed alteration of the $(\pi^0\pi^0)_S$ mass
spectrum in the $f_0(980)$ region with increasing $-t$ is satisfactorily
reproduced in the three variants of the proposed $\pi$ and $a_1$ exchange
model. In variant 1, this takes place with $A^2_\pi=340\times10^2$ (number of
events/GeV$^2$), $A^2_{a_1}=78.2$ (number of events/GeV$^2$), $C=-13.5$ GeV$^{-
2}$, and the slopes $b^0_\pi=9.4$ GeV$^{-2}$, $b^2_\pi=5.3$ GeV$^{-2}$, and $b^
0_{a_1}=5.4$ GeV$^{-2}$ which are rather typical for similar reactions (see the
solid curves in Figs. 1a-f). Note that the slope $b^2_\pi\approx5$ GeV$^{-2}$
had been observed in the reaction $\pi^+p\to\pi^+\pi^+n$ at $P^{\pi^-}_{lab}=
12.5$ GeV for the $\pi^+\pi^+$ production in the invariant mass region from
0.75 to 1.25 GeV [29]. In variant 2, the fit to the GAMS data is characterized
by the following values of the fitted parameters: $A^2_\pi=426\times10^2$
number of events/GeV$^2$), $A^2_{a_1}=639$ (number of events/GeV$^2$), $C=-4.4
$ GeV$^{-2}$, $b^0_\pi=12.4$ GeV$^{-2}$, $b^2_\pi=5.4$ GeV$^{-2}$, $b^0_{a_1
}=5.8$ GeV$^{-2}$, and $(g_{\sigma a_1\pi}\,g_{\sigma\pi\pi})/(g_{f_0a_1\pi}
\,g_{f_0\pi\pi})=0.16$ (see the dotted curves in Figs. 1a-f). In variant 3, the
fit gives $A^2_\pi=355\times10^2$ (number of events/GeV$^2$), $A^2_{a_1}=91.8$
(number of events/GeV$^2$), $C=-13$ GeV$^{-2}$, $b^0_\pi=10.1$ GeV$^{-2}$, $b^
2_\pi=5.2$ GeV$^{-2}$, $b^0_{a_1}=5.6$ GeV$^{-2}$, and $(g_{r_2a_1\pi}\,g_{r_2
\pi\pi})/(g_{r_1a_1\pi}\,g_{r_1\pi\pi})=-0.863$ (see the dashed curves in Figs.
1a-f). Note that in this case the $r_1$ and $r_2$ resonances interfere
destructively in the range $m_{r_1}<m<m_{r_2}$ in the $\pi^*\pi\to\pi\pi$
channel and constructively in the $a_1\pi\to\pi\pi$ channel.

Figure 4 shows the $t$ distributions of the $\pi^-p\to(\pi^0\pi^0)_S\,n$ events
for three $m$ regions $0.8-0.9$ GeV, $0.9-1$ GeV, and $1-1.1$ GeV which we
obtained for variant 1 using Eqs. $(1)-(4)$, and (13). The figure illustrates
how the one-pion exchange contribution falls and the $a_1$ exchange
becomes dominant in the $f_0(980)$ region as $-t$ increases.
Similar $t$ distributions take place also for variants 2 and 3.

Up to now we have adhered to the $t$ dependence factorization assumption.
However, it is easy to construct parametrizations which would permit one to
move beyond the scope of this assumption. A simplest example is provided by
variant 3 in which the amplitudes $T^0_0(m)$ and $R^0_0(m)$ are defined by
Eqs. (14) and (15) with $\delta_B=0$. For example, for the $\pi^*\pi\to\pi
\pi$ reaction amplitude $T^0_0(m,t)$, instead of Eq.(3) and Eq. (14) with
$\delta_B=0$, one can write a more general expression:
\begin{eqnarray} T^0_0(m,t)=\rho_{\pi\pi}\times\ \ \ \ \ \ \ \ \
\ \ \ \ \ \ \ \ \ \ \ \ \ \ \ \ \ \ \ \ \ \ \ \ \ \ \ \ \ \ \ \ \ \ \
\nonumber\\[3pt] \times\frac{g_{r_1\pi^*\pi}(t)[D_{r_2}(m)\,g_{r_1\pi\pi}+
\Pi_{r_1r_2}(m)\,g_{r_2\pi\pi}]+g_{r_2\pi^*\pi}(t)[D_{r_1}(m)\,g_{r_2
\pi\pi}+\Pi_{r_1r_2}(m)\,g_{r_1\pi\pi}]}{D_{r_1}(m)D_{r_2}(m)-\Pi^{\,2}_{r_1
r_2}(m)},\end{eqnarray} where the residues $g_{r_1\pi^*\pi}(t)$ and $g_{r_2\pi
^*\pi}(t)$ characterizing the $r_1$ and $r_2$ resonance production in the $\pi
^*\pi$ collisions, generally speaking, may be different functions of $t$ (at
$t=m^2_\pi,\ $ $g_{r_{1,2}\pi^*\pi}(m^2_\pi)=g_{r_{1,2}\pi\pi}$). Thus, if the
$t$ behaviors of these functions are appreciably different in a certain $t$
region, then it is natural that the $t$ dependence of the whole amplitude does
not
factorize in this region. However, we shall not exploit such a possibility,
first, because it requires incorporating at least two additional fitted
parameters (by one for every mechanism of the considered reaction), and
secondly, because a certain version of the extremal violation of
the factorization assumption has already been applied in Ref. [30] to explain
the GAMS data within the framework of the pure one-pion exchange model.
The results obtained in Ref. [30] are briefly discussed below.

\section{Comparison with the previous explanation}

As already mentioned 
in the Introduction, the explanation of the GAMS data on the reaction
$\pi^-p\to\pi^0\pi^0n$ [1] presented in Ref. [30] is based exclusively on the
one-pion exchange model (this immediately follows from Eqs. (2), (5), (6), Fig.
3a, and accompanying comments in Ref. [30] \footnote{It is worth noting that
the comment after Eq. (8) in Ref. [30] about a flat term which can effectively 
describe the contribution of the $a_1$ exchange to the $\pi N\to(\pi
\pi)_S\,N$ amplitude with the one-pion exchange quantum numbers from Eq. (5)
or Eq. (6) in Ref. [30] is misleading. In fact, at high energies, the $\pi$ and
$a_1$ Regge amplitudes have different spin structures and in the unpolarized 
cross section their contributions are noncoherent as already emphasized above.
So, the $a_1$ exchange has not been taken into account in Ref. [30] 
effectively.}\,). As a consequence of such a restriction, this explanation
leads to a strong violation of the $t$ dependence factorization assumption. We
can conveniently elucidate this assertion in terms of Eq. (16). Let us recall
that the authors of Ref. [30] used the $K$ matrix method to construct the $L=I=
0$ $\pi^*\pi\to\pi\pi$ reaction amplitude, and that, in the 1 GeV region
in the $K$ matrix, two resonances coupled to the $\pi\pi$ and $K\bar K$
channels and some background terms were taken into account. However, the
difference between the $K$ matrix representation for the amplitude
$T^0_0(m,t)$ obtained in Ref. [30] and Eq. (16) is unimportant to clear up the
question about the applicability of the pure one-pion exchange model for the
description of the GAMS data.

Thus, if one takes into account only the one-pion exchange mechanism for the
reaction $\pi^-p\to(\pi^0\pi^0)_S\,n$ and uses the parametrization with two
mixed
resonances coupled to the $\pi\pi$ and $K\bar K$ channels for the $L=I=0$ $\pi
^*\pi\to\pi\pi$ amplitude, then the observed alteration of the $(\pi^0\pi^0)_S$
mass spectrum can be understood only if the destructive interference between
two resonances at $m\approx1$ GeV, which occurs in the low $-t$ region, is
replaced by the constructive one with increasing $-t$. According to Eq. (16),
this means a change of the interference type between the terms proportional to
$g_{r_1\pi^*\pi}(t)$ and $g_{r_2\pi^*\pi}(t)$, which, in turn, is possible only
if, as $-t$ increases, one of
the residues, for example $g_{r_1\pi^*\pi}(t)$, decreases in absolute value,
vanishes at a certain value $t=t_0$, and then changes its sign. Also, this has
to occur at least for $-t<0.3$ GeV$^2$. Hence, according to such an
approach, the $t$ dependence of the amplitude $T^0_0(m,t)$ must not factorize
at $m\approx1$ GeV even in the low $-t$ region. In Ref. [30], the
following parametrization for the residues $g_{r_1\pi^*\pi}(t)$ and $g_{r_2\pi^
*\pi}(t)$ was postulated: \begin{eqnarray} g_{r_i\pi^*\pi}(t)=g_{r_i\pi\pi
}\,[\,1+\xi_i\,(1-t/m^2_\pi)\,t/m^2_\pi\,]\ , \qquad i=1,\,2\ . \end{eqnarray}
For the best fit $g_{r_1\pi\pi}=0.848$ GeV, $\xi_1=0.05
65$, $g_{r_2\pi\pi}=0.884$ GeV, and $\xi_2=-0.0293$ [30].
As is seen, the residue $g_{r_1\pi^*\pi}(t)$ vanishes at $t\approx-0.0728$ GeV$
^2$, and as $-t$ varies from 0 to 1 GeV$^2$, the functions $g^2_{r_1\pi^*\pi}(t
)$ and $g^2_{r_2\pi^*\pi}(t)$ increase, respectively, by approximately factors
of 22000 and 6000. In order to compensate this enormous rise, the authors
of Ref. [30] multiplied the $\pi^-p\to(\pi^0\pi^0)_S\,n$ one-pion exchange
amplitude by the overall form factor $F(t)=[(\Lambda-m^2_\pi)/(\Lambda-t)]^4$
with $\Lambda=0.1607$ GeV$^2$ which, however, they ascribed, for
unknown reasons, to the nucleon vertex \footnote{Note that this leads to
unsolvable difficulties. For example, if one describes
the well studied reaction $\pi^-p\to\rho^0n$ [6,42] using such a
form factor in the $\pi^*NN$ vertex it would be necessary to
introduce a $\pi^*\pi\,\rho$ residue which increases with $-t$. In turn,
this would lead to a rise of $d\sigma/dt$ for the process $\pi\pi\to\rho^0
\rho^0$. It is evident that such a picture is incompatible with conventional
ideas. Also, according to Eq. (17), we face a similar problem for the
reaction $\pi\pi\to(\pi\pi)_S(\pi\pi)_S$. Furthermore, the above form factor
would yield an abnormally sharp drop of the one-pion exchange contribution
to the differential cross section of the charge exchange reaction $pn\to np$.}.
As a result, they obtained formally a very good description of the GAMS data
on the $(\pi^0\pi^0)_S$ mass spectra. Recall that these spectra
($dN/dm$) correspond to the distribution $d^2N/dmdt$ integrated over $t$ in the
intervals indicated in Figs. 1a-f. Nevertheless, a detailed analysis shows that
the model of Ref. [30] predicts rather exotic $t$ distributions of the $\pi^-p
\to(\pi^0\pi^0)_S\,n$ events for $-t<0.2$ GeV$^2$. Figure 5 shows the
unnormalized $t$ distributions $(dN/dt)$ for three $m$ intervals $0.8<m<0.9$
GeV, $0.9<m<1$ GeV, and $1<m<1.1$ GeV which we obtained using the formulae
from Ref. [30]. The most discouraging feature of the presented picture is a dip
in $dN/dt$ whose location depends on $m$. In fact, this is a straightforward
consequence of a failure of the factorization for the amplitude $T^0_0(m,t)$.
The $t$ distribution for $0.8<m<0.9$
has a dip at $-t\approx0.1$ GeV$^2$ and, as is seen from Fig. 5,
changes very rapidly in the region $-t<0.2$ GeV$^2$. With increasing $m$, a dip
in $dN/dt$ moves to $t=0$. So, the $t$ distribution for $0.9<m<1$ GeV has a dip
at $-t\approx0.072$ GeV$^2$. For the mass interval $1<m<1.1$ GeV which already
belongs to the inelastic region of the reaction $\pi^*\pi\to\pi\pi$, a dip in
$dN/dt$ disappears. A comparison of the predictions for $dN/dt$ shown in Figs.
4 and 5 shows that the choice between our explanation of the GAMS data and
the explanation given by the authors of Ref. [30] can be easily realized
experimentally. To do this, it is sufficient to have data on $dN/dt$ in the
region $-t<0.2$ GeV$^2$ for the $m$ intervals $0.8<m<0.9$ GeV and $0.9<m<1
$ GeV. So far, however, neither the GAMS Collaboration [1] nor the E852
Collaboration [16] have published the data on the $t$ distributions.

Finally, let us emphasize that the best experimental test that we know of for
the $\pi^-p\to(\pi\pi)_S\,n$ reaction mechanisms are
measurements on polarized targets, because they will permit the interference to
be directly observed between the $\pi$ and $a_1$ exchange amplitudes.
As is known [24,25], in such experiments one can measure the triple
distribution (in $m$, $t$, and $\psi$) which at fixed $P^{\pi^-}_{lab}$ has the
form \begin{eqnarray} \frac{d^3N(\pi^-p_
\uparrow\to(\pi\pi)_S\,n)}{dmdtd\psi}=\frac{1}{2\pi}\,\frac{d^2N}{dmdt}\,+\,2\,
P\cos\psi \,I(m,t)\ ,\end{eqnarray} where $\psi$ is the angle between the
normal to the reaction plane and the (transverse) proton polarization $P$. The
first term in Eq. (18) corresponds to the distribution of events on an
unpolarized target. It can be presented as $(d^2N/dmdt)/2\pi=|M_{+-}^\pi(m,t)|
^2+|M_{++}^{a_1}(m,t)|^2$, where $M_{+-}^\pi(m,t)$ and $M_{++}^{a_1}(m,t)$ are
the $s$-channel helicity amplitudes with and without nucleon helicity flip,
due to the $\pi$ and $a_1$ exchange mechanisms, respectively. The second term
in Eq. (18) describes the nucleon polarization effects. The function $I(m,t)$
in this term is stipulated by the interference between the $\pi$ and $a_1$
exchange amplitudes and has the form: $I(m,t)=Im[\,M_{+-}^\pi(m,t)\,(M_{++}^
{a_1}(m,t))^*\,]$. In our model for the reaction
$\pi^-p\to(\pi^0\pi^0)_S\,n$ the amplitude $\sqrt{2\pi}\,M_{+-}^\pi(m,t)$ (and
respectively, $\sqrt{2\pi}\,M_{++}^{a_1}(m,t)$) is given by the expression
under the sign of modulus square in the first (second) term of Eq. (1). If one
neglects the $I=2$ $\pi\pi$ S-wave contribution, then the phase of the product
$M_{+-}^\pi(m,t)\,(M_{++}^{a_1}(m,t))^*$ in the elastic region (i.e. for $m<2m_
K$) would be completely defined by the Regge signature factors of the $M_{+-
}^\pi(m,t)$ and $M_{++}^{a_1}(m,t)$ amplitudes. With these provisos in mind,
one can easily write the function $I(m,t)$ in an explicit form for the three
considered variants. For example, for the most simple variant 1, up to a sign,
\begin{eqnarray} I(m,t)=\cos[\pi(\alpha_\pi(t)-\alpha_{a_1}(t))/2]\times \ \
\ \ \ \ \ \ \ \ \ \ \ \ \ \ \ \ \ \ \ \ \ \ \  \nonumber\\[3pt] \times\ \frac{
1}{2\pi}\ \left[A_\pi\ \frac{\sqrt{-t}}{t-m^2_\pi}\ e^{b^0_\pi(t-m^2_\pi)}
\ A_{a_1}\ (1+tC)\ e^{b^0_{a_1}t}\right]\left\{\sin(\delta^0_0)\,\frac{\sqrt{m
\Gamma_{f_0\pi\pi}(m)}}{|D_{f_0}(m)|}\right\}\ , \end{eqnarray} where, as
seen, the $t$ and $m$ dependences factorize. It is natural that the
pure one-pion exchange model [30] predicts $I(m,t)=0$.

\section {Conclusion}

We have suggested a new explanation of the GAMS results on the $f_0(980)$
production in the reaction $\pi^-p\to\pi^0\pi^0n$. A crucial role in our
explanation is assigned to the amplitude with quantum numbers of the $a_1$
Regge pole in the $t$ channel which is as of yet poorly studied. Moreover, we
consistently used the standard assumption of the $t$ dependence factorization.
On the other hand, if one attempts to explain the GAMS data in the framework of
the pure one-pion exchange model, as is done, for example, in Ref. [30], then
this assumption must be rejected from the outset. To test the correctness
of our explanation, the data on the $t$ distributions of the $\pi^-p\to(\pi^0
\pi^0)_S\ n$ events in the intervals $0.8<m<0.9$ GeV and $0.9<m
<1$ GeV, and the measurements of the reaction $\pi^-p\to(\pi^0\pi^0)_S\ n$
on polarized targets, which can clearly demonstrate the presence of the $a_1$
exchange mechanism, are needed.

Recently we have shown [43] that the new data on $d\sigma(\pi^-p\to a^0_0(980)n
)/dt$ can be explained within the framework of the Regge pole model only if the
reaction $\pi^-p\to a^0_0(980)n$ is dominated by the $\rho_2$ Regge pole whose
partner by exchange degeneracy is the $a_1$ Regge pole. To all appearance, the
time is right to study the pseudovector and pseudotensor Regge exchanges.
\vspace{0.2cm}

We would like to thank A.A. Kondashov and S.A. Sadovsky for supplying the GAMS
data in numerical form.
\vspace{0.2cm}

This work was partly supported by the INTAS Grant No. 94-3986.

\newpage

\begin{center}{\bf{Figure captions}}\end{center}

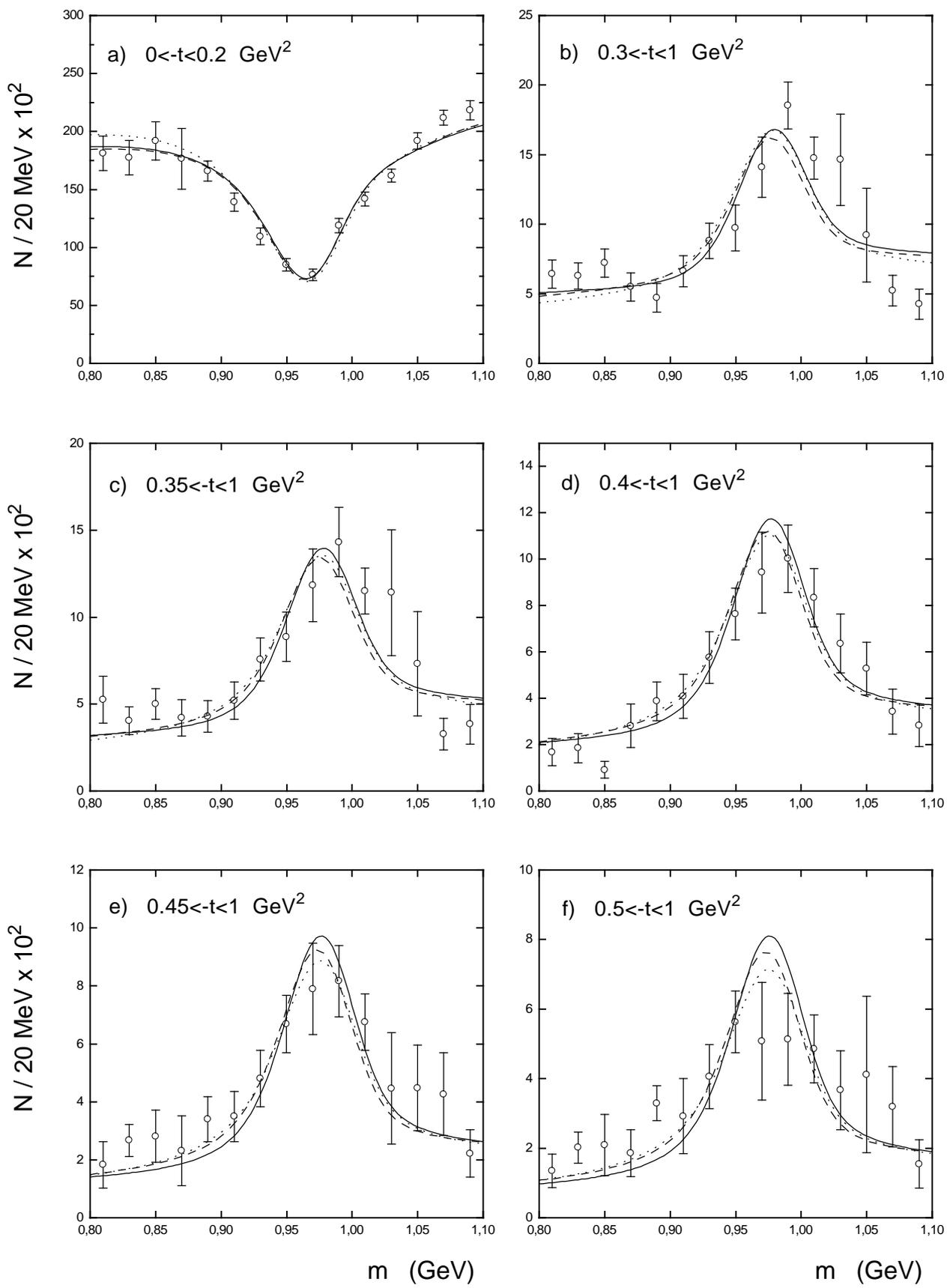
{\bf Fig. 1.} The S-wave $\pi^0\pi^0$ mass spectra in the reaction
$\pi^-p\to\pi^0\pi^0n$ for six $t$ intervals indicated in the figure.
The data were obtained by the GAMS Collaboration [1].
The curves correspond to the fits using the $\pi$ and $a_1$ exchange model
which is described in detail in the text. The solid curves correspond to
variant 1, the dotted curves to variant 2, and the dashed ones to variant 3.

\vspace{0.3cm}

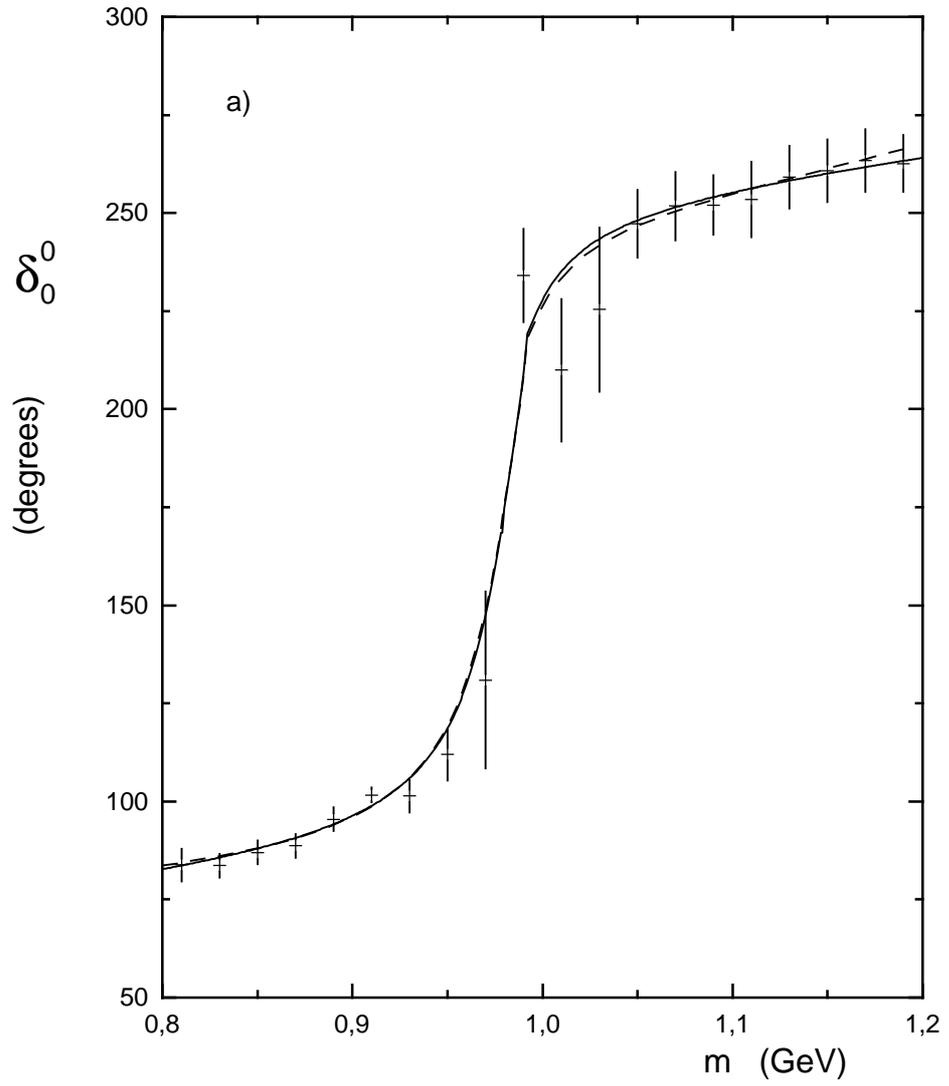
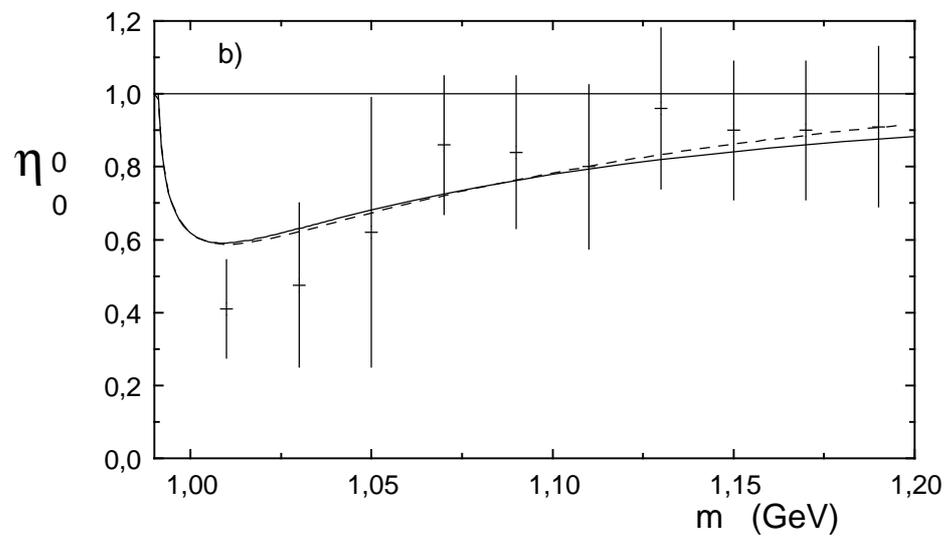
{\bf Fig. 2.} The phase shift $\delta^0_0$ (a) and the elasticity $\eta^0_0$
(b) pertaining to the $L=I=0$ $\pi\pi\to\pi\pi$ reaction amplitude $T^0_0(m)$
in the $f_0(980)$ region. The data are taken from Ref. [5]. The solid curves
correspond to the fit for variant 1 and the dashed curves to that for variant
3.

\vspace{0.3cm}

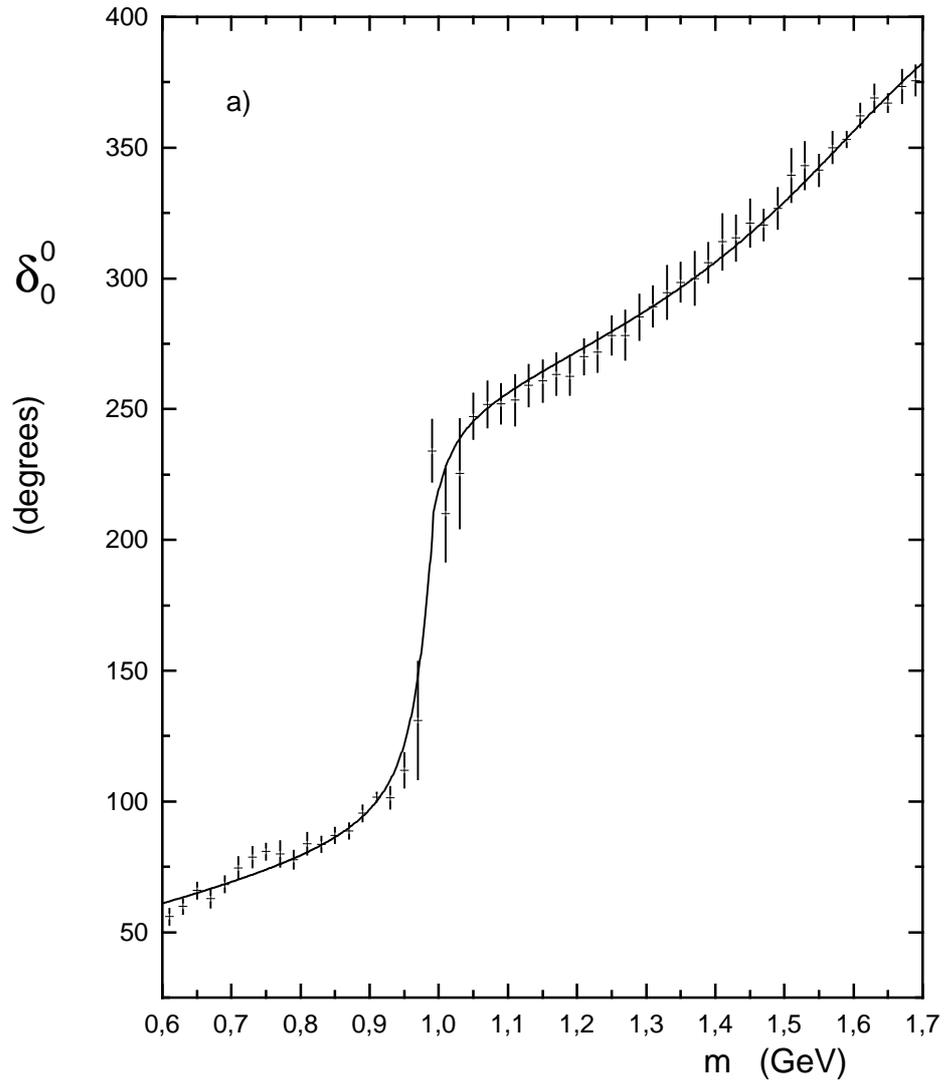
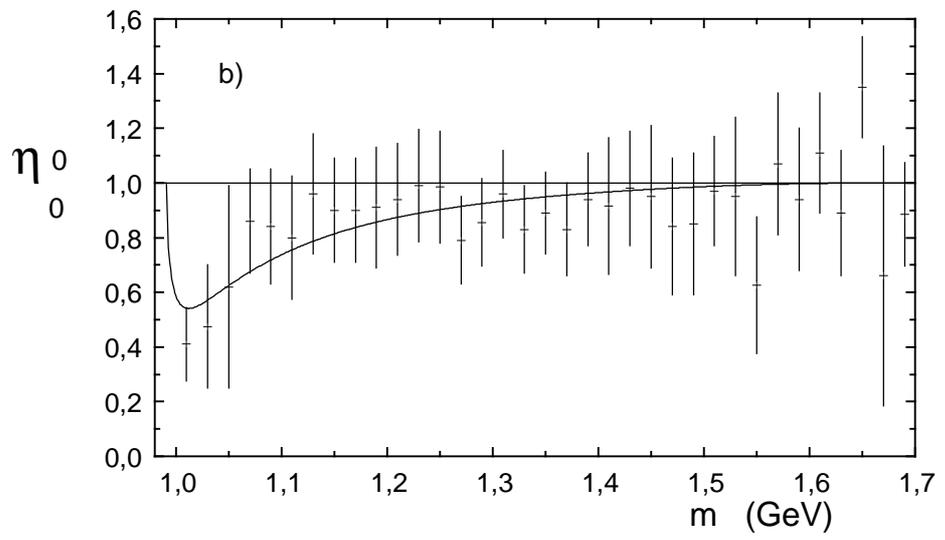
{\bf Fig. 3.} The phase shift $\delta^0_0$ (a) and the elasticity $\eta^0_0$
(b) pertaining to the $L=I=0$ $\pi\pi\to\pi\pi$ reaction amplitude. The data 
are taken from Ref. [5]. The curves correspond to the fit for variant 2.

\vspace{0.3cm}

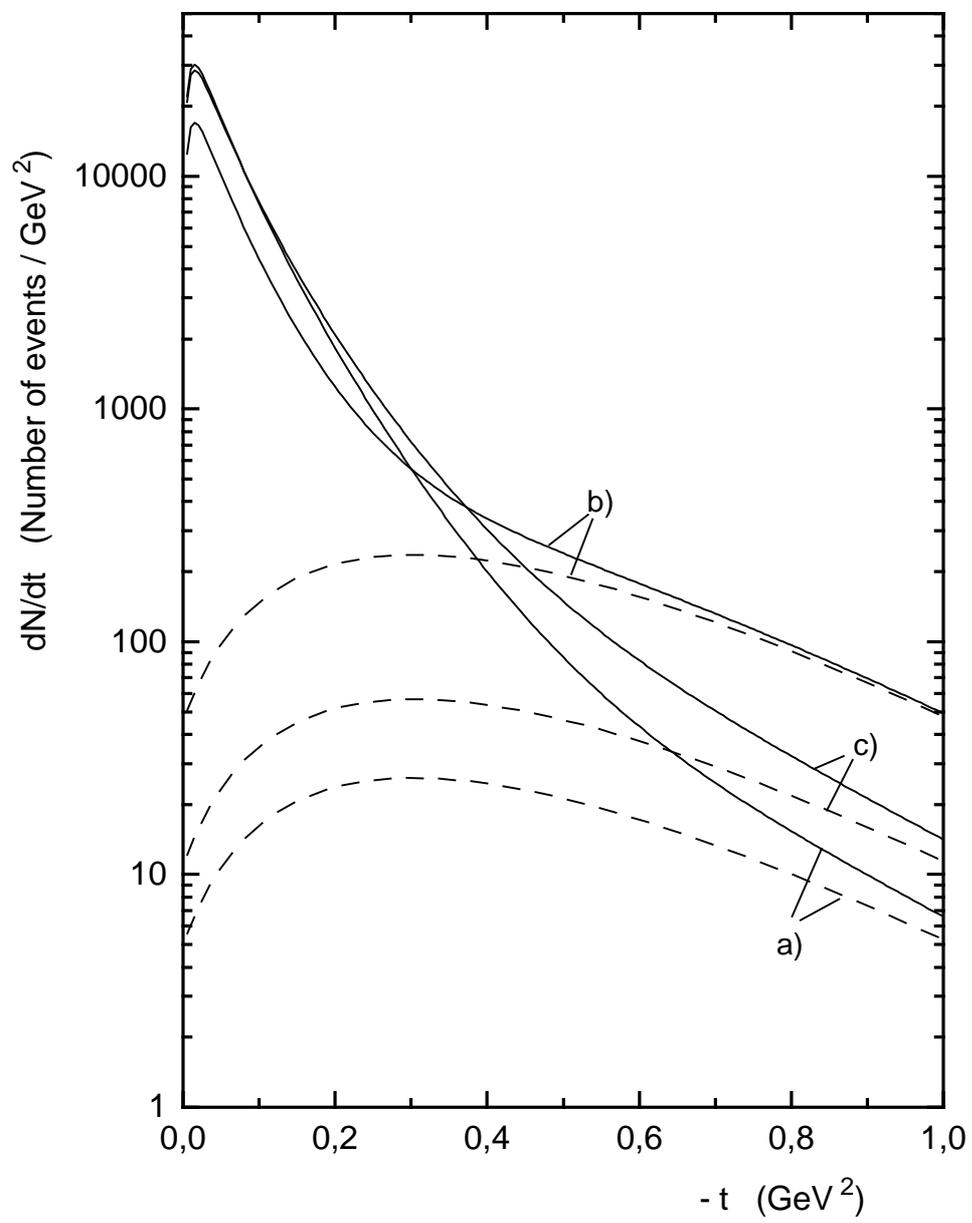
{\bf Fig. 4.} The $t$ distributions of the $\pi^-p\to(\pi^0\pi^0)_S\,n$ events
for three $m$ intervals a) $0.8-0.9$ GeV, b) $0.9-1$ GeV, and c) $1-1.1$ GeV
corresponding to variant 1. The solid curves correspond to the sum of the $\pi$
and $a_1$ exchange mechanisms and the dashed curves to the $a_1$ exchange
contribution.

\vspace{0.3cm}

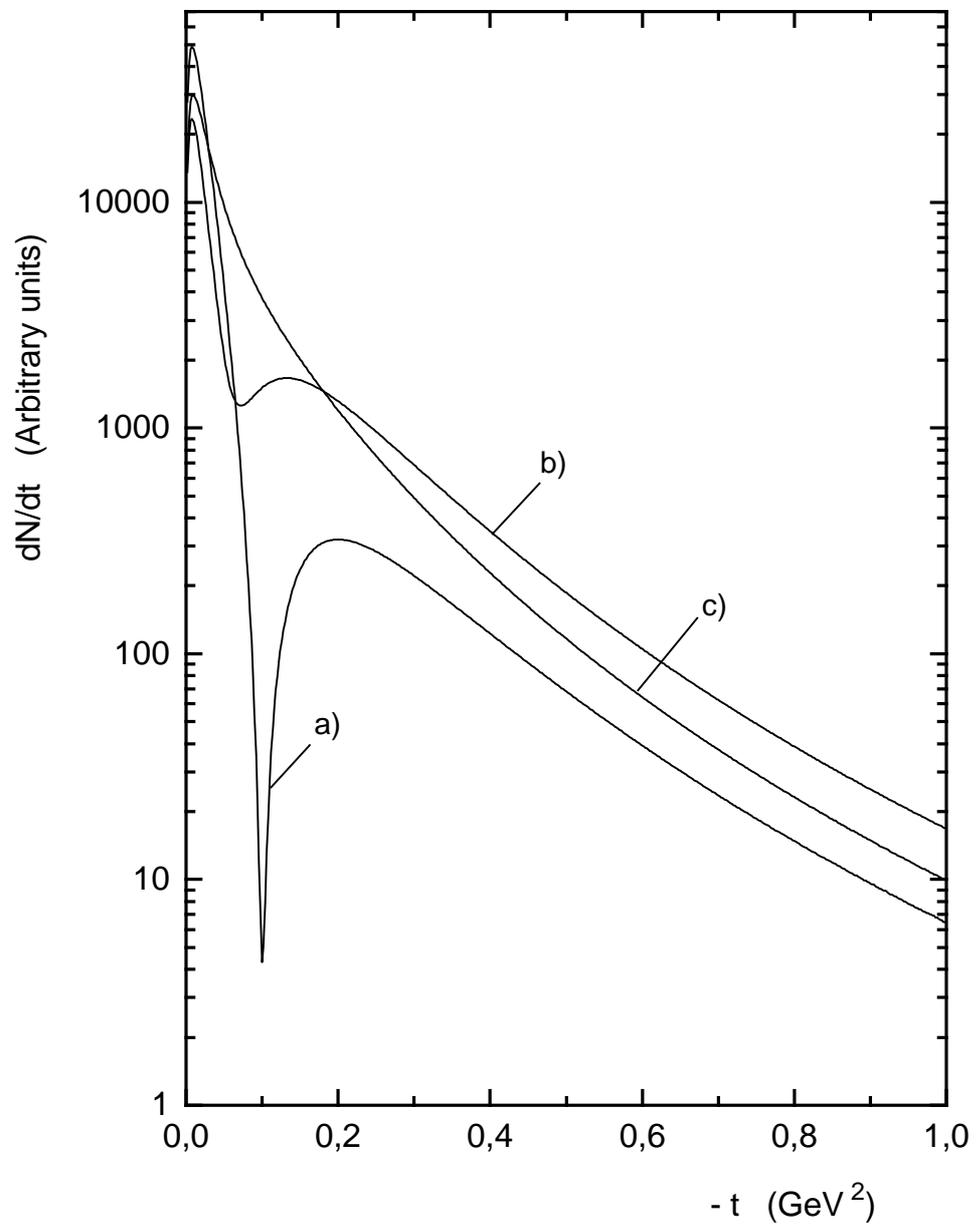
{\bf Fig. 5.} The unnormalized $t$ distributions for the reaction $\pi^-p\to(
\pi^0\pi^0)_S\,n$ for three $m$ intervals a) $0.8-0.9$ GeV, b) $0.9-1$ GeV, 
and c) $1-1.1$ GeV corresponding to the pure one-pion exchange model used in 
Ref. [30] (see text).

\end{document}